\documentclass{iau_FM}
\usepackage{graphicx}
\usepackage{tabularx}
\usepackage{soul}

\usepackage{rotating}
\usepackage{txfonts}
\usepackage{comment}
\usepackage{lscape}
\usepackage{threeparttable}
\usepackage{xcolor}
\usepackage{subfigure}
\usepackage{float}
\usepackage{ulem}

\newcommand{\HI}{H{\,\scriptsize I}}

\newcommand{\kms}{$\,$km$\,$s$^{-1}$}
\newcommand{\WHz}{$\,$W$\,$Hz$^{-1}$}
\newcommand{\ergs}{$\,$erg$\,$s$^{-1}$}

\newcommand{\msun}{{$M_\odot$}}
\newcommand{\msunyr}{{$M_\odot$ yr$^{-1}$}}

\newcommand{\tspin}{$T_{\rm spin}$}

\newcommand{\alwmath}[1]{\ifmmode#1\else$#1$\fi}

\def\emph#1{{\sl #1}}
\newcommand{\ltsima} {$\; \buildrel < \over \sim \;$}
\newcommand{\gtsima} {$\; \buildrel > \over \sim \;$}
\newcommand{\lta} {\lower.5ex\hbox{\ltsima}}
\newcommand{\gta} {\lower.5ex\hbox{\gtsima}}

\usepackage{txfonts}

\title[Associated HI absorption and 
AGN feedback] 
{Associated HI absorption and AGN feedback}

\author[Raffaella Morganti]   
{Raffaella Morganti$^{1,2}$
}

\affiliation{$^1$ASTRON, the Netherlands Institute for Radio Astronomy, Oude Hoogeveensedijk 4, 7991 PD, Dwingeloo, The Netherlands. 
\\[\affilskip]
$^2$Kapteyn Astronomical Institute, University of Groningen, Postbus 800,
9700 AV Groningen, The Netherlands\\email:{\tt morganti@astron.nl}}

\pubyear{2024}
\jname{Neutral Hydrogen in and around Galaxies in the SKA era} 
\editors{D.J. Pisano, Moses Mogotsi, Julia Healy \& Sarah Blyth, eds.}
\begin{document}

\maketitle

\begin{abstract}
The presence, distribution and kinematics of atomic neutral hydrogen in the central regions of galaxies can be traced by the \HI\ 21~cm line observed in absorption. Depending only on the strength of the radio continuum, the associated  absorption can trace the gas down to pc scale, which is ideal for exploring the \HI\ in the nuclear regions of radio AGN.  
This paper gives a brief overview of the main recent findings with particular focus on the AGN-driven \HI\ outflows. 
Absorption has made possible the discovery of fast and massive \HI\ outflows and their clumpy structure on pc scales. The similarities with the predictions of  numerical simulations confirm the impact of young radio jets in the feedback cycle and galaxy evolution. 
The field of \HI\ absorption is rapidly expanding thanks to new ``blind" surveys and the increased spectral capabilities of the radio telescopes. This opens many  possibilities for future discoveries and complements the studies of \HI\ emission.

\end{abstract}

\firstsection 
\section{Associated \HI\ absorption: relevance and properties}

Cold gas, including atomic neutral hydrogen (\HI), is a key ingredient for star formation and, as such, it plays an important role in galaxy evolution. On the other hand, the presence of an active black hole (i.e.\ an active galactic nucleus, AGN) and the release of energy associated to it, represents another key ingredient which can regulate the star formation, e.g.\ by driving gas outflows. Cold gas (including \HI) has been found to be the most massive component of AGN-driven outflows. Thus, this gas component is also involved in AGN (negative) feedback. 
All this highlights how valuable it is to have a tracer of the properties and role of cold gas in these competing processes. \HI\ absorption can provide such a tracer and this paper gives a short overview of the recent results and upcoming opportunities in the study of associated \HI\ absorption, i.e.\ absorption that traces the presence of atomic neutral hydrogen in and around the host galaxy of a radio source (see also \cite{Morganti18} for a more detailed review). 
Associated \HI\ 21-cm absorption provides at the same time information on the presence of \HI, allowing to push this search to relative high redshifts, and trace the interaction of radio jets (or radiation) with the interstellar medium (ISM) thanks to the signature of fast outflows that can be seen in the absorption profiles. 

The study of \HI\ absorption is now rapidly expanding thanks to the possibilities offered by the new ``blind" (or ``untargeted") surveys. Some of the findings summarised in this paper form the base and provide a reference point for these upcoming surveys. 
While this paper focuses on ``associated" \HI\ absorption, absorption can also trace intervening gas, i.e.\ \HI\ along the line-of-sight where a radio source is used only as background light. A description of the main goals and differences of these two approaches can be found in \cite{Kanekar04,Morganti15,Dutta22} and references therein. How to separate in an automatic way these two types of absorption can be sometime a challenge: the description of one proposed approach can be found in \cite{Curran21}.

Observations of \HI\ using associated absorption can reach similar column densities as obtained by the \HI\ emission observations ($\sim 10^{19}$ cm$^{-2}$) but can trace the gas on much smaller scales, depending on the distribution of the background radio continuum. Indeed, unlike \HI\ emission, the detection of absorption depends on the strength of the background radio continuum and, because of this, it allows to extend the search for \HI\ to high redshift, probe the evolution of the conditions of the ISM  and even trace energetic phenomena occurring on small (pc) scales, as long as enough continuum is available. 

The optical depth of the \HI\ cloud can  be derived from the spectrum using
\begin{equation}
\tau(V) = -\ln\big(1+\frac{\Delta T(V)}{c_{\rm f}T_c}\big) 
\end{equation}
where $T_c$ is the continuum flux, ${\Delta T(V)}$ the absorbed flux, and $c_f$ the covering factor. 
The column density of the \HI\ gas can be estimated as:
\begin{equation}
N_{\rm HI} = 1.82\times 10^{18} \  \frac{T_{\rm s}}{c_{\rm f}T_c} \int {|\Delta T(V)}|\ dV
\end{equation}
where  \tspin\ the spin temperature in K.

This formula shows that the column density depends also on two important parameters. One is the covering factor ($c_f$), which can be derived using observations at  high spatial resolution. More difficult is to determine the excitation temperature (\tspin) which usually is assumed to be 100 K (similar to the dominant component of \HI\ in our Milky Way), except for cases where the \HI\ shows extreme kinematics, where higher values (up to 1000 K) are used. 

\begin{figure}
\begin{center}
  \includegraphics[width=4.5in]{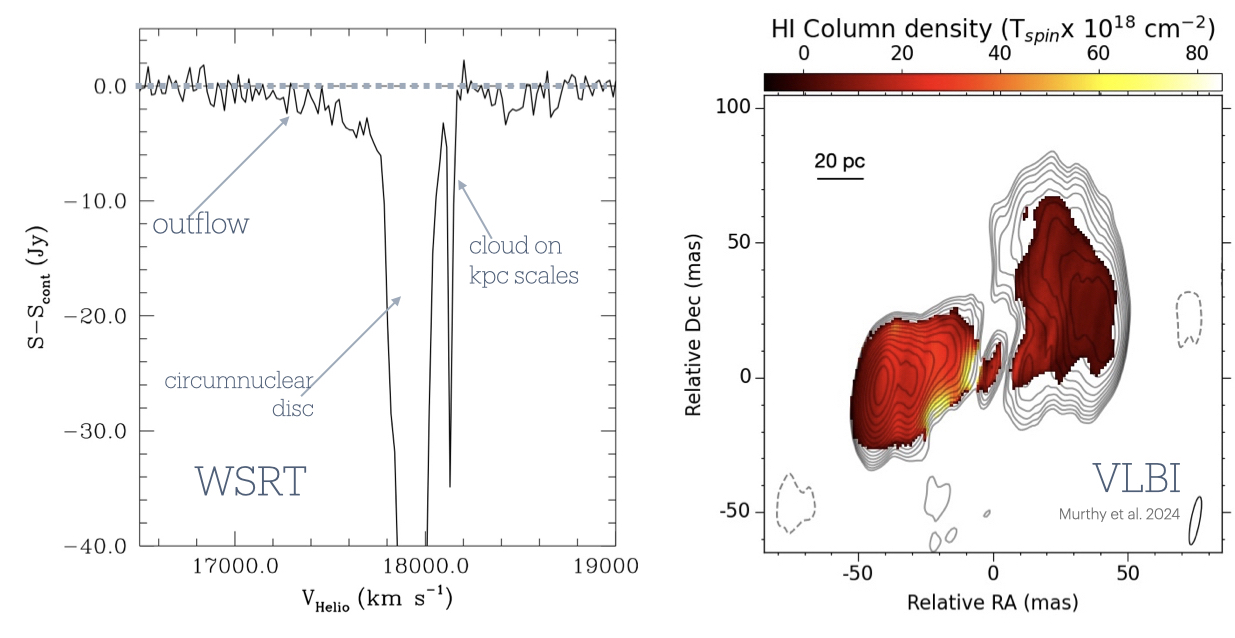}  
 \caption{Example of \HI\ absorption in the young radio source 4C~31.04, taken from \cite{Murthy24}. On the left is the integrated profile from the low resolution observations with the WSRT: a variety of structures (see labels) contribute to the global profile. On the right is the distribution of the \HI\ absorption using the VLBI where the various  components can be kinematically separated, see Sec. \ref{sec:outflows} for more details. }
   \label{fig:ExampleProfile}
\end{center}
\end{figure}

Although the presence of \HI\ absorption can be seen in a variety of galaxies, the requirement of having enough radio continuum flux on multiple scales, down to small (pc) regions, means that radio AGN tend to be more suitable targets for detecting \HI\ in absorption. 
Because these AGN (at least for the continuum fluxes searched for absorption) are preferentially hosted by early-type galaxies, this means that associated \HI\ absorption offers the opportunity to trace the presence of \HI\ in these types of host. 
Unsurprisingly, the main limitation of \HI\ observed - compared to emission - in absorption is that can only be traced against the underlying continuum.  
Only in an handful of nearby cases \HI\ both emission and absorption are observed, e.g.\ Cen A see \cite{Struve10}. 
On the other hand, the presence of absorption  provides a clear geometry for the gas (i.e.\ unambiguously located it in front of the continuum).  
Most of the available associated \HI\ absorption observations provide spatially unresolved absorption profiles, see Fig.\ \ref{fig:ExampleProfile} left. This is  because in the vast majority of the cases, \HI\ is detected in absorption against the central region of the radio source (see below). These profiles can, however, trace a variety of structures (see e.g.\ Fig.\,\ref{fig:ExampleProfile} left). Follow-up, high resolution observations (including VLBI) can be used to spatially resolve the distribution of the gas, as shown in Fig. \ref{fig:ExampleProfile} right (see also Sec.\ref{sec:outflows}) if enough radio continuum is present on those scales.

In summary, despite the limitations, the presence of associated \HI\ absorption is telling us about the presence of \HI\ even on small (nuclear) scales and up to large redshifts, much larger than what typically detected in emission. In this respect it is a powerful tracer of the evolution of \HI\ in galaxy and tracer of a variety of (even extreme) processes.

\subsection{From pointed observations to ``blind" surveys}

Most of what we know about associated \HI\ 21-cm absorption is the result of pointed observations. As reported by \cite{Aditya24}, about a thousand radio sources, covering a redshift range $0.1 < z < 5$,  have been targeted in searches for associated \HI\ absorption.

However, the field is rapidly changing thanks to the ``blind" surveys that are becoming available (see Yoon et al. and Gupta et al. these proceedings and references therein). These surveys take advantage of the wide-band correlators for radio telescopes which can detect \HI\ over a large range of redshifts. This is combined, in some telescopes, with the availability of a large field-of-view. Especially thanks to the wide-band, every radio continuum source in the field of view can be searched for \HI\ absorption (associated or intervening).

Particularly relevant are the dedicated \HI\ absorption surveys: 
\begin{itemize}
    \item ASKAP FLASH survey dedicated to \HI\ absorption in the redshift range $0.4 < z < 1.0$, see  \cite{Allison22,Yoon24}. The survey does an automatic search for \HI\ absorption (associated or intervening) against every continuum source without any optical pre-selection, see Yoon et al. these proceedings for details; 
    \item MeerKAT MALS: dedicated to trace intervening absorption against high-$z$ quasars, see \cite{Gupta16,Deka24a}, but can detect associated absorption in sources observed in the field of view, as shown in \cite{Deka24b} for the detection of \HI\ in a quasar at $z=1.3531$; 
    \item FAST drift scan surveys, FASHI, searching for \HI\ absorption in sources at $z<0.09$, \cite{Zhang24}, and $z<0.35$,\cite{Hu24}. 
\end{itemize}

In addition to these dedicated surveys, ``untargeted" searches for \HI\ absorption can in fact be done on any field observed with wide-band correlators and covering frequencies below 1420~MHz.

\begin{figure}
\begin{center}
  \includegraphics[width=3.8in]{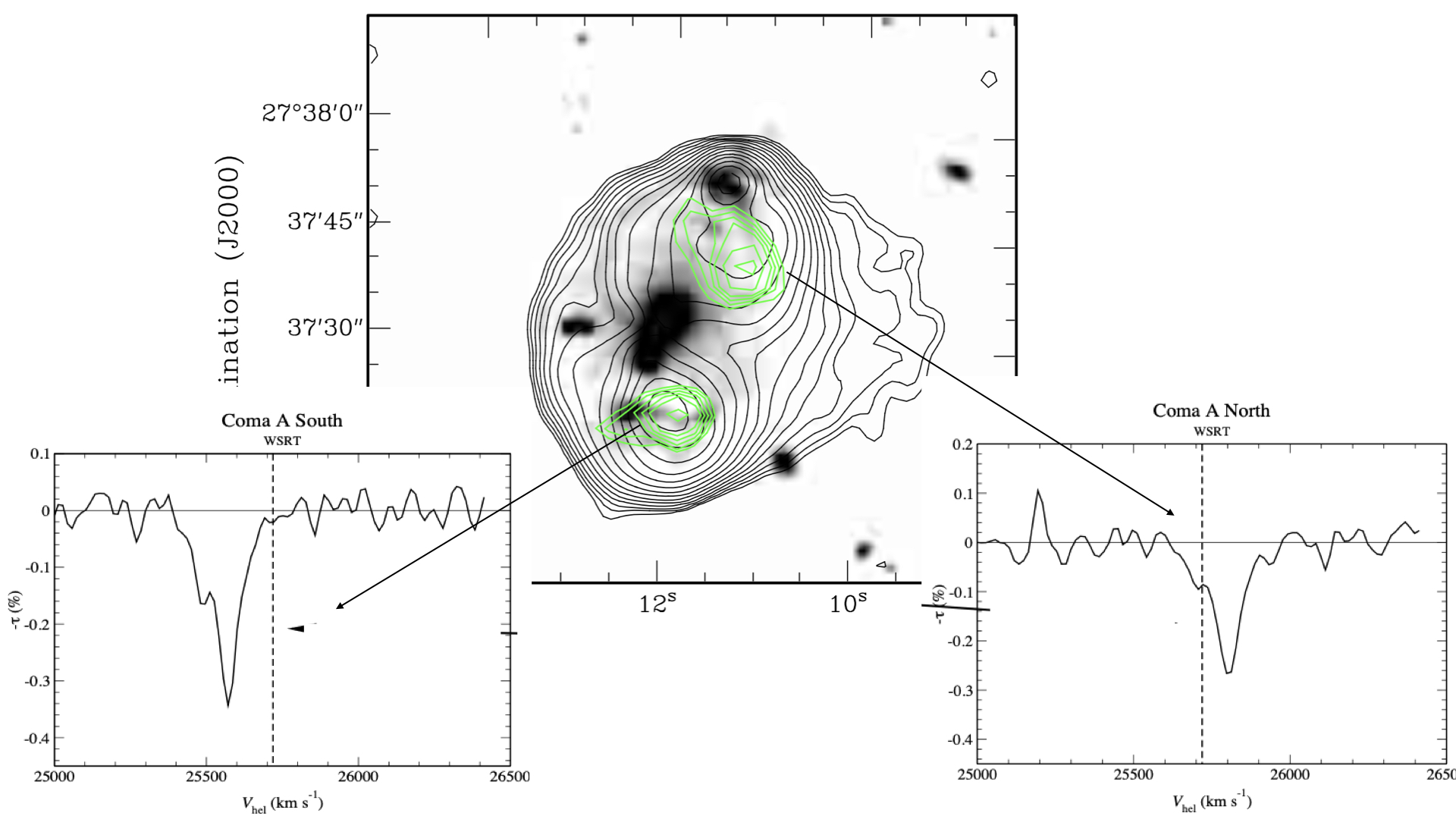}  
 \caption{ H$\alpha$ image (grey scale) with superimposed the contours of radio continuum  (black) and the contours of the column density of the \HI\ absorption,  in Coma~A, from \cite{Morganti02}. The velocities from the absorption profiles (redshifted and blueshifted with respect to the systemic velocity, marked by the dashed line) suggest the presence of a large, rotating \HI\ disc.}
   \label{fig:ComaA}
\end{center}
\end{figure}

\subsection{The structures traced by the associated \HI\ absorption}

Associated \HI\ absorption is mostly detected against the nuclear regions of galaxies (see below), but there are a few exceptions worth mentioning. For example, in the radio galaxy Coma~A redshifted and blueshifted absorption (with respect to the systemic velocity) have been detected against the radio lobes (up to $\sim 25$ kpc from the centre) suggesting the presence of a large \HI\ disc, see Fig. \ref{fig:ComaA} taken from \cite{Morganti02}.

Other cases of \HI\ absorption offset from the nuclear regions appear instead tracing the large-scale environment of the target galaxy. This is the case for 3C~433 where the gas is associated to a companion (with velocity only 50 \kms\ offset compared to the systemic velocity of 3C~433) located, in projection, against the southern radio lobe. Interestingly, this suggests that we may be witnessing an interaction between the radio lobe plasma and the companion,  see \cite{Murthy20}.
In other two known cases, PKS~0409-75 at z=0.693 found by ASKAP, \cite{Mahony22} and PKS~0405-12 at z=0.574 observed with MeerKat (Morganti et al. in prep), the difference between the velocity of the detected \HI\ and the systemic velocity of the targets is large (up to 3000 \kms\ for the former), suggesting that the \HI\ is tracing the large-scale environment of the radio sources. 
These cases show that, in order to do the correct optical identification and association of the \HI\ absorption with the radio continuum, high resolution observations for both the \HI\ and the radio continuum are needed as well as good ancillary data (e.g.\ optical images). This is particularly important for the upcoming ``blind" surveys. 
Situations like the one described (intermediate between associated and intervening cases), are likely to become more common with the expected increase of the number of \HI\ absorption. 
 
However, as mentioned above, associated \HI\ absorption is usually detected against the nuclear regions of galaxies and can be originated from various structures: i) (circum-nuclear) discs; ii) infalling clouds; iii) outflows/kinematically disturbed gas.
These structures can be co-exist and this can result in a complex absorption profile, see e.g.\ Fig. \ref{fig:ExampleProfile}. 

Except for cases of extreme kinematics of the gas, e.g.\ the extreme blueshifted absorption wings resulting from outflowing gas, see Sec. \ref{sec:outflows}, it is not always easy to identify which structure(s) describe the observed \HI\ integrated profile.
In some cases this task can be helped by a detailed (and high spatial resolution) knowledge of the radio continuum. An illustration of this is the case of the famous radio galaxy 3C~84 in the Perseus cluster where only thanks to the detection of the pc-scale radio continuum counterpart of the circumnuclear disc, the \HI\ absorption structure could be explained, see \cite{Morganti23} for details.
Indeed, because it is the morphology of the radio continuum which defines what part of the \HI\ distribution is detected, the shape of the resulting \HI\ absorption profile depends on that. This means that even the absorption from a regularly rotating disc can result in an asymmetric \HI\ profile if the background continuum has a complex/asymmetric morphology. 
Examples of how the high resolution continuum information can be used to predict the \HI\ absorption are presented in e.g.\ \cite{Murthy21} and  \cite{Maccagni23}. The latter shows how, in the case of NGC~3100, this method has allowed the authors to identify, in addition to the regularly rotating gas,  a redshifted component of \HI\ likely fuelling the AGN, see \cite{Maccagni23} for details.

\begin{figure}
\begin{center}
  \includegraphics[width=2.4in]{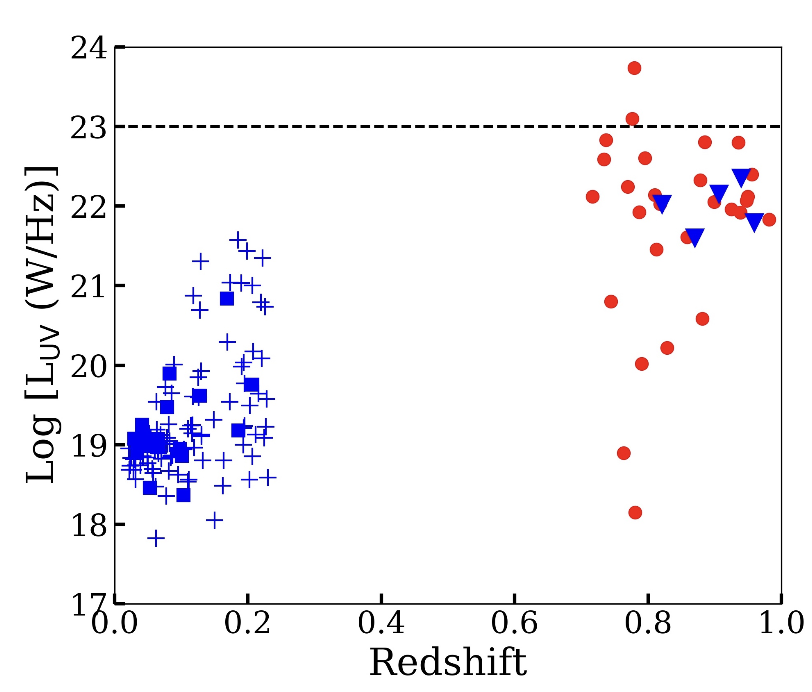}  
 \caption{Comparison of the UV luminosities of the sources in \cite{Murthy22} (red circles), the extended radio sources in the sample of \cite{Aditya19} (blue triangles), and the extended sources in \cite{Maccagni17} (blue squares: detections; blue crosses: non-detections). The dashed line marks the threshold UV luminosity L$_{UV} = 10^{23}$ \WHz, above which it has been argued that the radiation from the AGN ionises the \HI\ in the host galaxy \cite{Curran13}. Only a minority of high-$z$ sources below this limit are detected, suggesting that the UV luminosity is not the only relevant parameter, from \cite{Murthy22}.}
   \label{fig:UV}
\end{center}
\end{figure}

\subsection{Associated \HI\ absorption at low and high redshift}

As mentioned above, what we know about the occurrence and properties of associated \HI\ absorption come mostly from pointed observations carried out at low and high redshift. 
\cite{Aditya24} reported that about thousand radio sources have been targeted in searches for associated \HI\ absorption, with about 200 detections which are mostly at $z<0.4$. Only eleven detections have been reported to date at $z > 1$. 
Comparing the properties and the detection rate of \HI\ absorption in low and high-$z$ can be used to probe differences in the conditions of the gas. 
As described below, the presence of differences (in particular in the detection rate) is still matter of debate and study and it will be one of the major goals for the new surveys.

At low redshift (i.e.\ $z\lesssim 0.4$) a relatively consistent picture, albeit with a number of still open questions, is emerging as illustrated by, among others, \cite{Maccagni17,Murthy21,Chandola13,Gupta06,Chandola24}, using a variety of telescopes (e.g.\ VLA, WSRT, GMRT and FAST). The majority of the \HI\ absorption detections have peak optical depth ranging between $\tau_{peak} \sim 0.008$ and $\tau_{peak} \sim 0.15$, see e.g.\ the distributions presented in \cite{Maccagni17} and \cite{Curran21}.  
The dearth of detections with lower optical depth is likely due to the sensitivity of the available observations.  
Cases with high optical depth (i.e.\ $\tau_{peak} > 20$\%) have been found but are rare. The typical column densities of the \HI\ are in the range $10^{20}$ to few $\times 10^{21}$  cm$^{-2}$ (for \tspin = 100 K). For these low-$z$ sources, the detection rates does not show a dependence on redshift or on radio luminosity, but should be noted that most of the sources are well below luminosities of $ \sim 10^{26}$\WHz, see e.g.\ \cite{Maccagni17,Chandola24}.

Most interestingly, peaked spectrum sources, i.e.\ sources which are considered to have young radio jets, see \cite{ODea21} for a review, have been found to have higher detection rates ($\sim 25$\%) of \HI\ absorption, see e.g.\ \cite{Gupta06,Maccagni17,Glowacki17,Chandola24}. 
These newly born radio sources represent a particularly interesting group of objects especially for their potential impact on the host galaxy, see Sec. \ref{sec:outflows}.

\begin{figure}[t]
\begin{center}
  \includegraphics[width=5.4in]{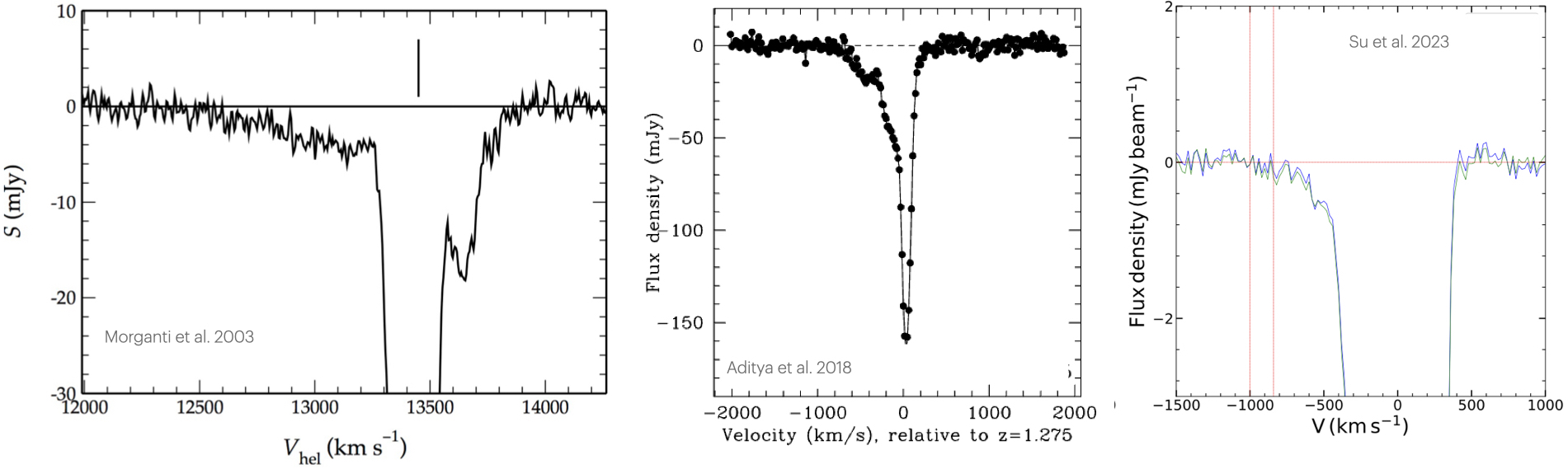}  
 \caption{Examples of \HI\ absorption with blueshifted wing indicating the presence of outflowing gas. See also Fig. \ref{fig:ExampleProfile} for an other example. The profiles (from left to right) are from 3C~293 observed with the WSRT 
 \cite{Morganti05} which represent one of the first case of \HI\ outflows, TXS 1245-197 observed with GMRT \cite{Aditya18a} and SDSS J145239.38+062738.0 observed with FAST \cite{Su23}. }
   \label{fig:OutflowProfiles}
\end{center}
\end{figure}

In extended sources the detection rate of \HI\ absorption is found to be lower ($\sim 10$\%) than in peaked sources. This can have interesting physical implications, e.g.\ about the impact of the radio jet on the cold gas. However, no study has yet been able to securely rule out an effect of covering factor. The lower detection rate could be the result of the relative weakness of the radio cores in extended radio sources: the depth of the available observations would then not be enough to reach the optical depth needed for detecting \HI. 
This can only be established by high resolution continuum observations quantifying the flux of the radio continuum in the central region and the required sensitivity of the \HI\ observations.  

Finally, the occurrence of \HI\ absorption gives a lower limit of the presence of \HI\ in early-type galaxies, as discussed in \cite{Maccagni17}. The detection rates discussed above are, to first order, similar to what found for \HI\ emission for early type galaxies, see \cite{Serra12}. No detection of \HI\ absorption has been yet found via stacking experiments of the undetected sources. This is likely due to the still limited number of targets that can be used for the stacking, combined with the limited sensitivity of the stacked observations, see \cite{Maccagni17} for details. This is hopefully going to change with the new ``blind" surveys. 

Because of the relevance for tracing evolution in the conditions of the neutral gas, many studies have been dedicated to the search of \HI\ absorption at higher redshift.
Surprisingly, the situation about the detection rate of associated \HI\ absorption at high redshift (i.e.\ $z >0.4$) sources is still not completely settled. One reason for this is the quality of the observations at the lower frequencies (for the redshifted \HI), which are often badly affected by terrestrial radio interference. This should hopefully improve with the new surveys (and SKA) which are operated from radio-quiet zones. 

In general, the detection rate appears much lower for radio sources at high-$z$, see e.g.\ \cite{Curran13,Dutta22,Aditya24} and refs therein. 
However, similar to what seen at low redshift, the detection rate appears also to depend on the morphology of the sources, with the vast majority of the \HI\ detections at high-$z$ found in peaked sources, see e.g.\ \cite{Glowacki19,Aditya18a,Aditya24}.
Compact but flat spectrum quasars have a low detection rate, \cite{Aditya18b}.
The low detection rate of \HI\ absorption at high-$z$ is interesting because it could indicate changes in conditions of the gas but they can also be the result of bias affecting the observed samples. In particular, the sources observed at high redshift are all high power radio sources (with a radio luminosity $\gtrsim 10^{27}$ \WHz, see \cite{Maccagni17}), therefore different from the low-$z$ samples. These powerful sources, that would mostly be classified as Fanaroff-Riley type 2, tend to have relatively faint cores which could make the detection of \HI\ absorption (for an optical depth similar to what found at low-$z$) challenging.  Indeed, the presence of biases is illustrated by the finding of two high-opacity \HI\ 21 cm absorbers at $z\sim 1.1$ from an untargeted search  of the DEEP2 fields (\cite{Chowdhury20}). Being low-luminosity AGN, these objects would not have been targeted by pointed observations. 

However, other effects can influence and reduce the detection rate. For example, the effect of high UV luminosity characterising most of the high-$z$ sources has been suggested by \cite{Curran13} to play a role, because the strong radiation would ionise the surrounding gas. 
Similarly, the high radio luminosity could affect the \tspin\ of the gas close to the core of the radio sources. 
The effect of the UV luminosity has been questioned by the results of \cite{Aditya19}, which have reported the detection of \HI\ absorption in sources with high UV luminosity, and \cite{Murthy22}. The latter by observing  extended sources with relatively low UV luminosity (at least lower than the $10^{23}$ \ergs), have confirmed the low detection rate, see Fig. \ref{fig:UV}.

Thus, the possibility of explaining the low detection rate as a result of the evolution in the general conditions of the high-$z$ medium is still open, see \cite{Murthy22}. This would suggest that high-$z$ gas is characterised by higher spin temperature, as also observed for the high-$z$ damped Lyman-$\alpha$ absorbers (e.g.\, \cite{Kanekar14}), or by a lower column density.
Clarifying the occurrence and properties of the \HI\ in high-$z$ galaxies (taking into account possible observational and selection biases) is one of the main tasks for the upcoming \HI\ absorption surveys.

\begin{figure}
\begin{center}
  \includegraphics[width=4.4in]{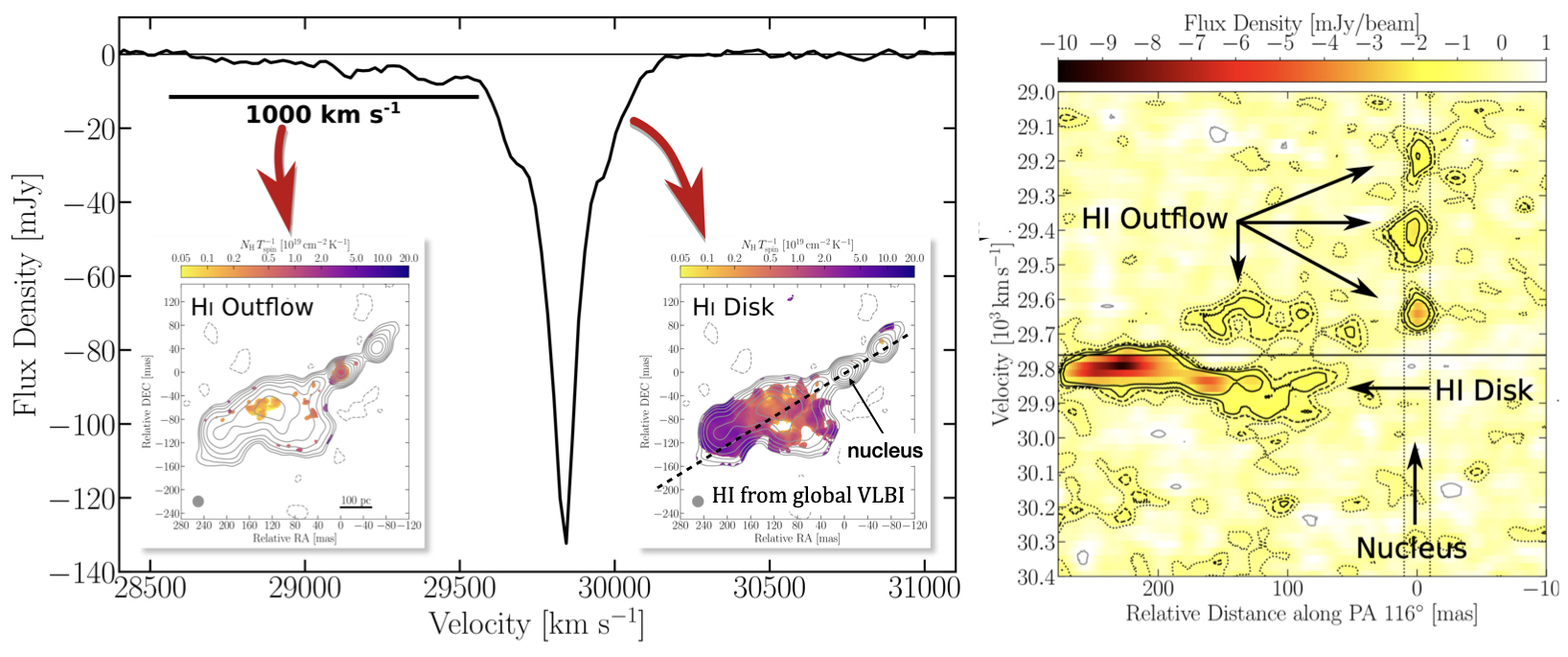}  
 \caption{The case of 3C~236: integrated profile from the WSRT (left) and images from the global VLBI observations where the absorption associated with a circumnuclear disc and with the outflow can be separated (insets), \cite{Schulz18}. The figure on the right shows a position-velocity plot taken along the dashed line. This shows that the outflowing gas is mostly organised in clouds, see \cite{Schulz18}. 
 }
   \label{fig:VLBIoutflow}
\end{center}
\end{figure}

\section{\HI\ absorption and AGN feedback: fast outflows}
\label{sec:outflows}

As can be seen in Fig.\ \ref{fig:ExampleProfile} and Fig.\ \ref{fig:OutflowProfiles}, a group of radio AGN show \HI\ absorption profiles characterised by extreme blueshifted wings. The radio galaxy 3C~293, see Fig. \ref{fig:OutflowProfiles} (left) taken from 
\cite{Morganti05}, has been one of the first objects where this extreme kinematic of the \HI\ has been observed. The large blueshifted velocities  unambiguously indicate the presence of fast outflowing neutral atomic hydrogen. While outflow in AGN have been routinely found in warm ionised gas, the presence of a cold component was unexpected. 

Interestingly, similar extreme velocities have been subsequently found in cold molecular gas (traced by CO)  confirming that AGN-driven outflows, despite being the results of highly energetic phenomena,  include an unexpected component of cold gas. Outflows of cold molecular gas are now successfully traced (also in emission) in many objects using ALMA and NOEMA. The presence of the cold  component of gas is likely the result of the fast cooling of  the shocked gas. 
Most importantly, the cold gas (\HI\ and cold molecular gas) has been found to be most massive component of the AGN-driven outflows. The mass outflow rates found for \HI\ outflows reaches up to $\sim 40$ \msunyr. This highlights the relevance of these gas outflows for feedback.

The \HI\ outflows are typically identified by shallow ($\tau<<1$\%) blueshifted wings, which are often found in addition to the more common rotating structures (circumnuclear discs) as in the examples  shown in Figs. \ref{fig:ExampleProfile} and \ref{fig:OutflowProfiles}. 
These shallow and broad wings are challenging to detected. Not only deep observations are needed but also, and most importantly, a bandpass, stable down to 10$^{-3}$ - 10$^{-4}$. 
The typical column density found in \HI\ outflows, is a few $\times 10^{21}$ cm$^{-2}$ for \tspin = 1000 K, see \cite{Schulz21}. 

The mechanism driving these outflows (jets or radiation) can be different for different objects.
However, it is interesting to note that the vast majority of the fast and massive \HI\ outflows are found in young (or restarted) radio sources (with ages $<10^6$ yr). Thus, this suggests a role for the radio jet. This role  appears to be more prominent in the first phase of the expansion of the radio plasma in the surrounding ISM. Interestingly, this is consistent with the predictions from numerical simulations (see below).

Identify the location and distribution of the outflowing \HI\ is needed for understanding its origin and for the comparison with the simulations.  High resolution (VLBI) observations have helped in this by allowing to trace the \HI\ down to pc-scale.
Figure \ref{fig:VLBIoutflow} shows the case of 3C~236 where the \HI\ is partly distributed in a disc aligned with the dust-lane of the host galaxy and partly in clouds outflowing at $\sim 600$ \kms. These clouds can be seen already in the inner few $\times 10$ pc from the nucleus, and they  have an average density of  $n_e \sim 120$ cm$^{-3}$ and mass around  few $\times 10^4$\msun, see \cite{Schulz18} for details and \cite{Schulz21} for other examples. 
Because in some cases only a fraction of the \HI\ outflow is recovered by the VLBI, it is likely that also a diffuse component of outflowing gas may be present at larger scales.

\begin{figure}
\begin{center}
  \includegraphics[width=4.0in]{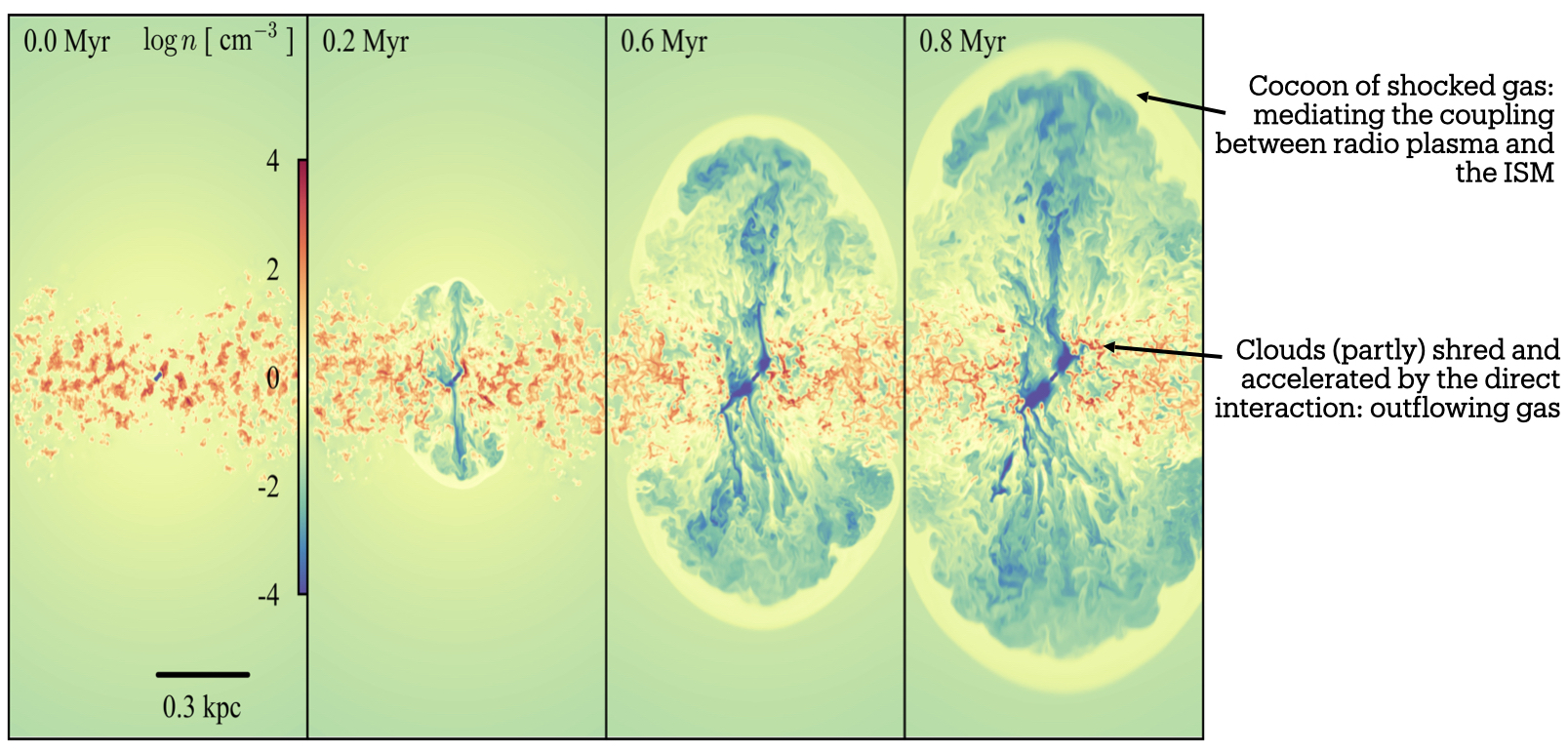}  
 \caption{Numerical simulation showing different evolutionary stages of a radio jet expanding in a clumpy medium, from \cite{Mukherjee18} and \cite{Murthy22}.
 }
   \label{fig:Simulations}
\end{center}
\end{figure}

These results are relevant because they can be directly compared with numerical simulations of an evolving young jet expanding in the surrounding medium. According to simulations, see  e.g.\ \cite{Mukherjee18} and refs therein, a jet expanding in a clumpy medium (as the one revealed by the VLBI observations) can couple more efficiently with the gas.
As illustrated in Fig. \ref{fig:Simulations}, the impact of such coupling is seen both in outflowing gas as well as in the creation of a cocoon of shocked gas which can expand and affect, over a broad angle, the galactic ISM. 
Thus, the impact of the radio jets may not be only in driving gas outflows but also in increasing the velocity dispersion as result of the expanding cocoon. Indeed, while the mass outflow rates for \HI\ outflows are significant (see above), the outflows have been typically found spatially limited to the inner kpc of the galaxy. Thus, on the larger scales the increasing of velocity dispersion of the gas as effect of the expanding cocoon can be a more subtle but effective signature of the impact of the radio plasma. 

\begin{figure}
\begin{center}
  \includegraphics[width=2in]{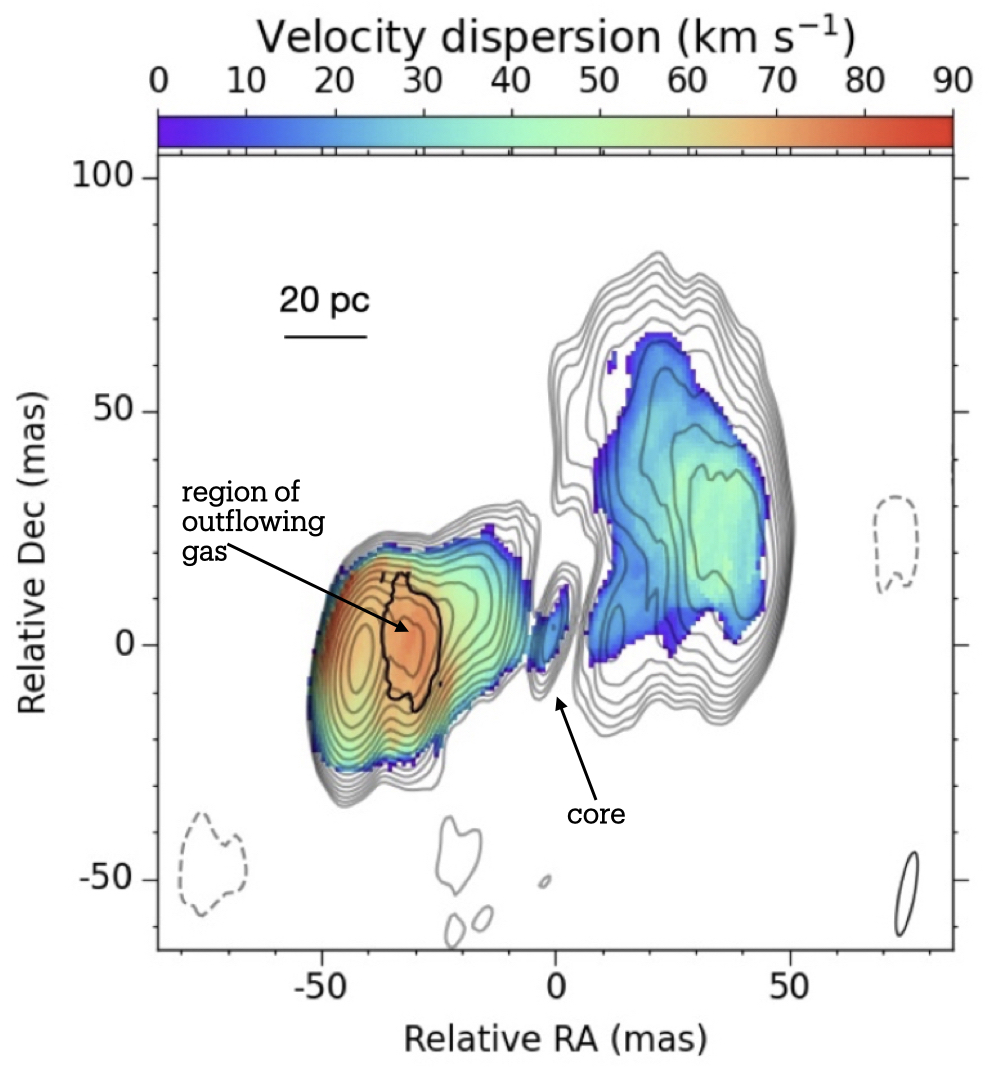}  
 \caption{The velocity dispersion on VLBI scale of the HI in the young radio galaxy 4C31.04, taken from \cite{Murthy24}. The region (about 35 pc in size) with thick contours include the outflowing \HI\ seen in the global profile shown in Fig. \ref{fig:ExampleProfile}.}
   \label{fig:fig4Cdispersion}
\end{center}
\end{figure}

This is seen for example in the case of 4C~31.04, the young radio galaxy where, as shown in Fig. \ref{fig:ExampleProfile}, in addition to a circumnuclear disc an \HI\ outflow has been detected and followed-up with VLBI observations, see \cite{Murthy24} for details. 
Figure \ref{fig:fig4Cdispersion} shows the velocity dispersion of the \HI\ on pc scales. The gas outflow (of 1.4 \msunyr) is seen located in a limited region at $\sim 35$pc from core (thick black contour). However, the entire distribution of the \HI\ is characterised by gas with high velocity dispersion ($\geq 40$ \kms).
Thus, the kinematics of the \HI\ in this object has been explained as the result of jets/lobes expanding in the gas disc and disturbing the kinematics and the properties of the gas producing outflowing gas where a direct interaction jet-cloud is occurring, and injecting turbulence resulting in high velocity dispersion of the gas in the rest of the gas.
In summary, the general agreement between the predictions of the simulations and the results from the \HI\ absorption observations (and more recently also confirmed by observations of the cold molecular gas with ALMA and NOEMA) confirms that radio jets can substantially affect the surrounding medium and, therefore, have a role in the feedback process and evolution of the host galaxy. 

As final remark, it is also interesting to investigate the 
survival of atomic gas in subkpc-scale outflows. The recent study of \cite{Perucho24} has shown that shocks driven by low-power radio jets ($\leq 10^{43}$ \ergs) are not strong enough to ionise entire atomic gas clouds. Instead, for higher jet power  the gas has to be sufficiently dense for the cooling times to be very short and the recombination rate to be high, see \cite{Perucho24} for details.

\section{Summary and future}
The results described above show that associated \HI\ absorption can be a powerful tracer of the presence, properties and kinematics of atomic neutral hydrogen in radio sources. It also gives the possibility of pushing the search for this gas to redshifts higher  than what the \HI\ emission can do. The upcoming ``blind" surveys will expand the number of known associated \HI\ absorption and will improve the statistics particularly needed for high-$z$ sources. 

The synergy of the \HI\ absorption and the cold molecular gas for understanding the properties of the AGN-driven outflows has been already shown in a number of studies. Furthermore, it is also interesting to have an independent tracer of the atomic neutral gas, like NaI D absorption, to compare with the properties of the outflows from the \HI\ absorption. Interestingly, \cite{Lehnert11} found the occurrence and properties of outflows derived from the NaI D absorption in radio sources to first order similar to what derived from \HI\ absorption. The NaI D as tracer has now become  even more promising with the advent of JWST, which allows to extend the study of neutral gas using NaI D to high-$z$. Using this tracer \cite{Belli24} have found the dominance of the outflow of neutral gas in a galaxy at redshift 2.45, with a hydrogen column density they derive ($9.6\times 10^{20}$ cm$^{-2}$)  to first order similar to what found in the \HI\ outflows described above.  This confirms the relevance of this phase of the gas for feedback. Hopefully the complementary of the two tracers will be exploited more in the future.

The results above also show the strong synergy between \HI\ absorption and the study of the radio continuum and cold molecular gas. The knowledge of the radio continuum emission down the small (pc) scale is essential for better interpreting the observed \HI\ absorption profiles and for pushing the search for \HI\ also on pc scales using the VLBI.
The synergy with the studies of the cold molecular gas will allow in the future to use the \HI\ detected from large surveys to identify interesting objects for e.g.\ ALMA follow up where complementary information (including e.g.\ high resolution CO emission) can be derived to obtain a complete view of the complex processes ongoing in the central regions of radio AGN.

\acknowledgements
I would like to thank the SOC of the conference "Neutral Hydrogen in and around Galaxies in the SKA Era" for the invitation. I would also like to thanks my collaborators, and in particular Tom Oosterloo and Suma Murthy: without their contribution some of the results presented in this overview would not have been obtained.

\end{document}